# Emergency Equity: Access and Emergency Medical Services in San Francisco

Corresponding Author: Robert Newton[a], 310 Leonhard Building, The Pennsylvania State University, University Park, PA, USA, ran5267@psu.edu

Soundar Kumara[b], 310 Leonhard Building, The Pennsylvania State University, University Park, PA, USA, u1o@psu.edu

Paul Griffin[c], 310 Leonhard Building, The Pennsylvania State University, University Park, PA, USA, pmg14@psu.edu

[a] PhD Student, Harold and Inge Marcus Department of Industrial and Manufacturing Engineering

[b] Allen E. Pearce and Allen M. Pearce Professor of Industrial Engineering, Harold and Inge Marcus Department of Industrial and Manufacturing Engineering

[c] Professor of Industrial Engineering, Harold and Inge Marcus Department of Industrial and Manufacturing Engineering

**Abstract –**

In 2020, California required San Francisco to consider equity in access to resources such as housing, transportation, and emergency services as it re-opens its economy post-pandemic.

Using a public dataset maintained by the San Francisco Fire Department of every call received related to emergency response from January 2003 to April 2021, we calculated the response times and distances to the closest of 48 fire stations and 14 local emergency rooms. We used logistic regression to determine the probability of meeting the averages of response time, distance from a fire station, and distance to an emergency room based on the median income bracket of a ZIP code based on IRS statement of income data.

ZIP codes in the lowest bracket ($25,000-$50,000 annually) consistently had the lowest probability of meeting average response metrics. This was most notable for distances to emergency rooms, where calls from ZIP codes in the lowest income bracket had an 11.5% chance of being within the city's average distance (1 mile) of an emergency room, while the next lowest probability (for the income bracket of $100,000-$200,000 annually) was 75.9%. As San Francisco considers equity as a part of California's "Blueprint for a Safer Economy," it should evaluate the distribution of access to emergency services.

## 1. Introduction

In 2020, California included an equity focus as part of its "Blueprint for a Safer Economy" plan to reopen its economy. The plan required each county with a population greater than 106,000 to meet an equity metric as well as submit a targeted equity investment plan (California Department of Public Health, 2021). The equity metric—from the Healthy Places Index—captured areas including economics, housing, transportation, and healthcare. The only indicator used for access to healthcare was the number of insured adults. (Delaney, et al., 2018) As San Francisco considers equity in its community as part of this "Blueprint," we considered the problem of equitable access to emergency services across its communities.

Municipalities establish fire departments to protect citizens and property from fires. Firefighters are often also trained to perform emergency medical services and search and rescue (Murray, 2013). Emergency rooms play a critical role in community healthcare, as they provide continuous access to care regardless of time or day, and increased time waiting for the interventions at an emergency department is closely tied with increased mortality (Claret, et al., 2016). As distance and time are correlated in the case of medical emergencies, distance from care is also positively correlated with mortality (WIlde, 2013). While cities set their own standards for response times, there is evidence that emergency medical service (EMS) responses under five minutes reduce mortality risk (Blackwell & Kaufman, 2002).

In 1997, The Paramedic Division of the San Francisco Department of Public Health merged with the Fire Department, enlisting and certifying city firefighters to handle life basic and advanced or life-threatening EMS calls. (Lian, et al., 2019) This paper capitalizes on the dense, diverse, urban environment of San Francisco coupled with its rich publicly available data to estimate the geographic distribution of its emergency services.

Many have proposed models to optimally site fire stations around a given city. The spatial nature of these models often applies set radii around each of a set of fire stations to ensure adequate protection and service to a population. (Murray, 2013) Some models provide a weighted distance to a station's coverage informed by the risk of fire in an area. (Yao, et al., 2019) The discussion of how to calculate fire risk has developed from predicting future emergencies using past incidents to the classification of the kinds of fire based on the property types or land uses based on colocation quotients. (Xia, et al., 2019) In all of these cases, time or distance are acceptable definitions of service radius.

To address the effects of suburban sprawl on a fire department, Lambert, et al. (Lambert, et al., 2012) used median income in a ZIP code as a predictor of mean response time to an emergency. With the goal of reduced mortality in San Francisco, a dense city with limited parking and notorious traffic, Lian et al. (Lian, et al., 2019) estimated San Francisco Fire Department (SFFD) response times based on call type, the time the call was received, and the ZIP code of the call in their regression model. Our work differs from previous work that focused on equitable outcomes or efficiency (Enayati, et al., 2019). In particular, we examine emergency responses in San Francisco using the median income of its ZIP codes as a predictor to estimate equitable access to services.

## 2. Methods

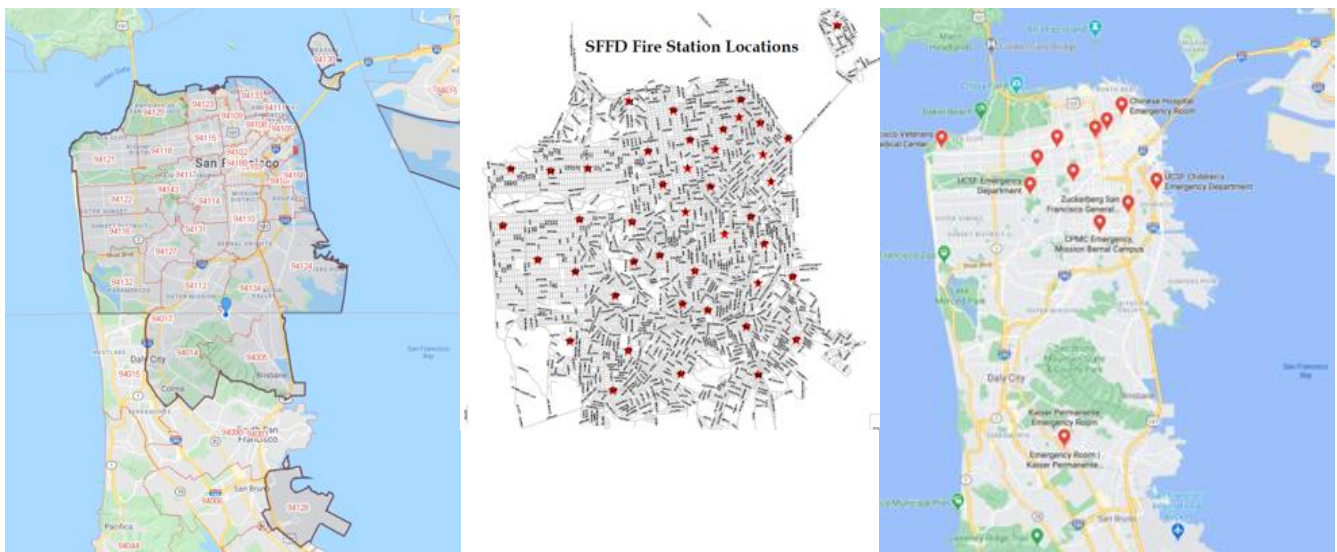

Figure 1 - (left to right) (a) ZIP codes in the SFFD dataset, (b) SFFD station locations, (c) San Francisco emergency rooms.

### a. Data Description

SFFD emergency call data reported all emergency calls over the period of 01 January 2003 through 26 April 2021—a total of 556,369 unique calls. (Anon., 2020) These data include alarm times, arrival times of SFFD personnel, location of each call (both as an address with ZIP code and as latitude and longitude coordinates), the station responsible for the area of each call, the primary situation, and actions taken. The dataset noted 39 unique five-digit ZIP codes (as shown in Figure 1a) including the San Francisco International Airport. We removed incidents where the SFFD action were to only investigate, inform, or standby, and calls that were canceled enroute, which led to a final dataset of 227,265 calls.

As of May 2021, San Francisco reported 48 fire stations, including three companies at the San Francisco International Airport. We derived latitude and longitude of each fire station (shown in Figure 1b (San Francisco Fire Department, n.d.)) from their street addresses with an online tool at Get-Direction.com.

Additionally, we determined the addresses of the 14 major emergency rooms around San Francisco from an area search in Google Maps (shown in Figure 1c). We likewise converted these addresses to latitude and longitude.

Tax Year 2018 ZIP Code Data from the Internal Revenue Service (IRS) included data on the number and adjusted gross income of tax filings in each ZIP code. (Internal Revenue Service, n.d.) The IRS had tax data for 37 of the 39 ZIP codes and classified each tax filing into one of six brackets: i) under $25,000, ii) $25,000 - $50,000, iii) $50,000 - $75,000, iv) $75,000 - $100,000, v) $100,000 - $200,000, and vi) $200,000 or more. 2010 US Census data included population and household information for each of the 39 ZIP codes (U.S. Census Bureau, 2010).

### b. Model Discussion

We considered two metrics for SFFD emergency responses: i) how far an emergency call is from a station (miles), and ii) how quickly the SFFD responded to a call (minutes). We measured response times from alarm—when the 911 operator dispatched the first unit to a call—to arrival—when the first unit arrived to the scene (Eq. 1). We used great circle adjustment to convert latitudes and longitudes into angular distance (radians) (Eq. 2) (Williams, n.d.), and earth's radius to convert to miles (Eq. 3) (JPL Solar System Dynamics, n.d.) to calculate Euclidean distances for each call from all 48 stations and cataloged the station with the minimum distance. We used the same methodology to measure distance from each call to the nearest emergency room.

$$Response\ Time = Arrive\ Time - Alarm\ Time \quad \text{(Eq. 1)}$$

$$Distance(radians) = 2asin\sqrt{\left(\sin\left(\frac{lat_{call}-lat_{stn}}{2}\right)\right)^2 + \cos(lat_{call})\cos(lat_{stn})\left(\sin\left(\frac{long_{call}-long_{stn}}{2}\right)\right)^2} \quad \text{(Eq. 2)}$$

$$Distance(miles) = Distance(radians) \times 3959 \quad \text{(Eq. 3)}$$

We based the threshold for the logistic regression on the mean of each metric and used the median income brackets of ZIP codes as predictors using the logit model (Eq. 4). We used all types of calls from the reduced dataset described for distances and times from fire stations, but used only EMS-specific calls for emergency room distances—a total of 64,415 events.

$$ln(odds) = \beta_0 + \beta_1 x_1 + \beta_2 x_2 + \beta_3 x_3 + \beta_4 x_4 \quad \text{(Eq. 4)}$$

where

$$x_1 = \begin{cases} 1 & if\ ZIP\ median\ income\ is\ \$25,000 - \$50,000 \\ 0 & otherwise \end{cases}$$

$$x_2 = \begin{cases} 1 & if\ ZIP\ median\ income\ is\ \$50,000 - \$75,000 \\ 0 & otherwise \end{cases}$$

$$x_3 = \begin{cases} 1 & if\ ZIP\ median\ income\ is\ \$75,000 - \$100,000 \\ 0 & otherwise \end{cases}$$

$$x_4 = \begin{cases} 1 & if\ ZIP\ median\ income\ is\ \$100,000 - \$200,000 \\ 0 & otherwise \end{cases}$$

### 3. Results and Analysis

, while 15.14% were rescue and emergency medical calls. We calculated average response times and distances and, from these averages, used 5 minutes, 0.4 miles, and 1.0 mile as thresholds in the logistic regression for response time, distance to station, and distance to emergency room, respectively. The histograms in Figure 2 graphically depict the values shown in Table 2. The histogram of distances from calls to an emergency room, Figure 2c, shows a bimodal distribution with maxima at approximately 0.5 and 2.7 miles.

Table 1 shows that only 12.04% of calls were in response to a fire, while 15.14% were rescue and emergency medical calls. We calculated average response times and distances and, from these averages, used 5 minutes, 0.4 miles, and 1.0 mile as thresholds in the logistic regression for response time, distance to station, and distance to emergency room, respectively. The histograms in Figure 2 graphically depict the values shown in Table 2. The histogram of distances from calls to an emergency room, Figure 2c, shows a bimodal distribution with maxima at approximately 0.5 and 2.7 miles.

Table 1. Call categories and their frequencies from the original dataset (Note: False alarms were removed from the dataset prior to logistic regression to simplify the analysis).

| Category | Incidents | Percentage |
|---|---|---|
| False Alarms | 251445 | 45.19% |
| Rescue/EMS | 84209 | 15.14% |
| Service Calls | 79138 | 14.22% |
| Fire | 67550 | 12.14% |
| Good Intent Calls | 33470 | 6.02% |
| Hazardous Condition | 31773 | 5.71% |
| Special Incident | 6170 | 1.11% |
| Rupture | 2113 | 0.38% |
| Severe Weather/Natural Disaster | 355 | 0.06% |

Table 2. Summary statistics from SFFD responses from 01 January 2003 through 26 April 2021.

| | Average | Standard Error | Range |
|---|---|---|---|

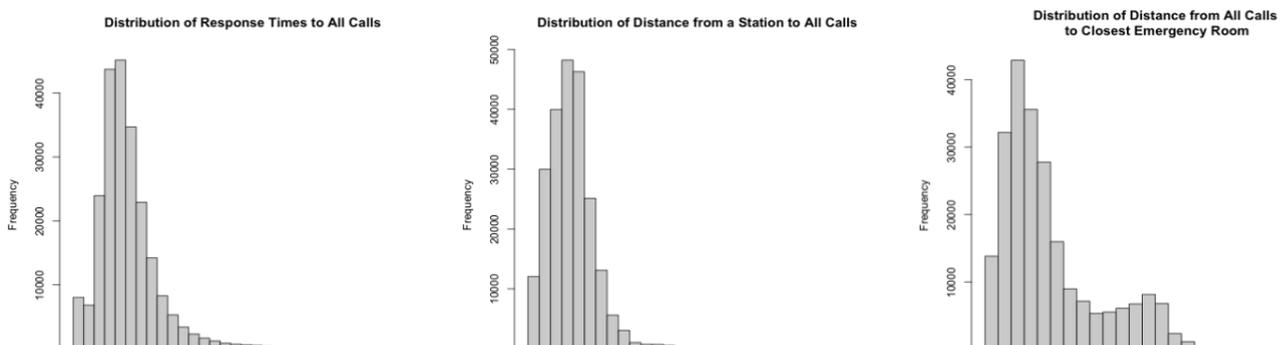

Figure 2 - (left to right) (a) Histogram of SFFD response times, (b) Histogram of distances from SFFD stations, (c) Histogram of distances from emergency rooms.

| | | | |
|---|---|---|---|
| Response Time from Station | 327 seconds | 1.13 seconds | 0 – 83515 seconds |
| Distance from Station to Call | 0.38 miles | 0.00043 miles | 0.00084 – 4.41 miles |
| Distance from Calls to Emergency Room | 0.97 miles | 0.0016 miles | 0.0015 – 5.09 miles |

Logistic regression yielded probabilities for each group of ZIP codes associated with a median income bracket as shown in (Eq. 5), (Eq. 6), and (Eq. 7). When evaluated with the Hosmer-Lemeshow Goodness of Fit Test, all three models had a *p*-value of 1.0 and $\chi^2$ values of $2.53 \times 10^{-21}$, $1.26 \times 10^{-19}$, and $1.07 \times 10^{-19}$, respectively, suggesting a good fit of the logistic regression model.

$$Pr(Response\ Time \leq 5\ minutes) = \frac{1}{1+e^{-(0.213-0.238x_1+0.0249x_2+0.292x_3+0x_4)}} \tag{Eq. 5}$$

$$Pr(Station\ Distance \leq 0.4\ miles) = \frac{1}{1+e^{-(-0.224+0x_1+0.365x_2+1.176x_3+0.706x_4)}} \tag{Eq. 6}$$

$$Pr(ER\ Distance \leq 1.14\ miles) = \frac{1}{1+e^{-(0.913-3.314x_1+0x_2+1.645x_3+0.356x_4)}} \tag{Eq. 7}$$

Table 3 shows differences between call response times and distances to ZIP codes in the lowest median income bracket and the rest of San Francisco, as these low income ZIP codes are in the only bracket with probability less than 0.50.

Table 3. Probabilities from logistic regression for time and distance from fire stations.

| Median Income | Number of ZIP Codes | Incidents | Probability Response ≤ 5 min | Probability Distance ≤ 0.4 mi | Population |
|---|---|---|---|---|---|
| $25,000 - $50,000 | 6 | 35377 | 0.494 | 0.444 | 245040 |
| $50,000 - $75,000 | 15 | 92407 | 0.559 | 0.535 | 474200 |
| $75,000 - $100,000 | 7 | 47936 | 0.624 | 0.722 | 181140 |
| $100,000 - $200,000 | 9 | 48966 | 0.553 | 0.618 | 146250 |

The results of the logistic regression regarding the distance of EMS calls to an emergency room in Table 4 show that while fewer EMS calls originated from the six ZIP codes in the lowest median income bracket, the probability those calls were within the average distance to an emergency room was less than 10%. The next lowest probability of 71.4% was for ZIP codes in the next highest bracket. As shown in Figure 3, lower income ZIP codes tended toward the southeast part of the city as well as Treasure Island in the northeast.

Table 4. Probabilities from logistic regression for distance from an emergency room.

| Median Income Bracket | Number of ZIP Codes | Incidents | Probability Distance to Emergency Room ≤ 1 mi |
|---|---|---|---|
| $25,000 - $50,000 | 6 | 10235 | 0.0979 |
| $50,000 - $75,000 | 15 | 26612 | 0.714 |
| $75,000 - $100,000 | 7 | 13644 | 0.928 |
| $100,000 - $200,000 | 9 | 13924 | 0.780 |

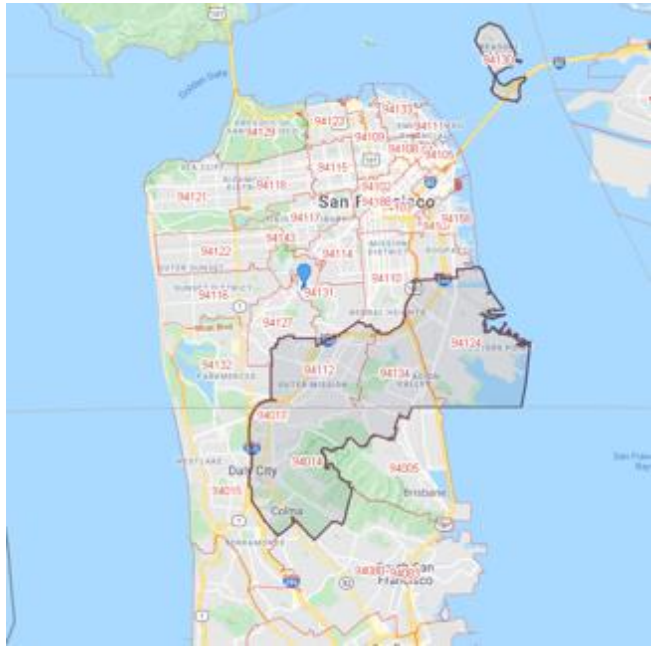

Figure 3 - ZIP codes in the $25k-$50k bracket (lowest median income bracket in the dataset and had the lowest probabilities of meeting the thresholds in the three logistic regression models)

The average distances and response times for each ZIP code in the region are provided in Appendix A. None of the ZIP codes in the two highest brackets had an average response time more than 8 minutes, while two ZIP codes in the lowest bracket had an average response time greater than 11 minutes. Additionally, there was no ZIP code in the lowest bracket with an average distance to an emergency room less than 1.6 miles.

With IRS data specific to 2018, we also ran the logistic regression using only the SFFD calls from 2018, with 15,115 events.

Table 5 shows an improved average response time by approximately 30 seconds, but still close to five minutes, with insignificant changes in the average distances to fire stations or emergency rooms. Consequently, we used the same thresholds (5 minutes, 0.4 miles, and 1 mile for response time, distance from fire station, and distance to emergency room, respectively) for analysis of 2018 data.

Table 5. Summary statistics from SFFD 2018 responses.

| Description | Average | Standard Error | Range |
|---|---|---|---|
| Response Time from Station | 296 seconds | 1.71 seconds | 0 – 5577 seconds |
| Distance from Station to Call | 0.38 miles | 0.0017 miles | 0.00085 – 4.41 miles |
| Distance from Calls to Emergency Room | 0.96 miles | 0.0059 miles | 0.0015 – 5.09 miles |

Table 6 shows improved probabilities for response time and distance from a fire station across all income brackets, while Figure 4a and 4b show a similar distribution in 2018 as across all years. The lowest income bracket remained the only group of ZIP codes with less than 50% probability of being within 0.4 miles from a fire station and maintained the lowest probability of response time less than 5 minutes. Both response time and distance models had Hosmer-Lemeshow $p$-values of 1.0 and $\chi^2$ of $5.28 \times 10^{-21}$ and $5.00 \times 10^{-21}$, respectively, indicating a good fit.

Table 6. Probabilities from logistic regression for 2018 time and distance from fire stations.

| Median Income Bracket | Number of ZIP Codes | Incidents | Probability Response ≤ 5 min | Probability Distance ≤ 0.4 mi | Population |
|---|---|---|---|---|---|
| $25,000 - $50,000 | 6 | 2257 | 0.571 | 0.482 | 245040 |
| $50,000 - $75,000 | 15 | 6719 | 0.660 | 0.542 | 474200 |
| $75,000 - $100,000 | 7 | 3053 | 0.748 | 0.744 | 181140 |
| $100,000 - $200,000 | 9 | 3076 | 0.603 | 0.582 | 146250 |

Table 7, however, shows minimal improvement for the lowest income bracket on access to emergency rooms and Figure 4c shows a similar distribution of distances from calls to emergency rooms. Again, the Hosmer-Lemeshow test indicated a good fit with a *p*-value of 1.0 and $\chi^2$ of 6.44x10⁻¹³.

Table 7. Probabilities from logistic regression for 2018 distance from an emergency room.

| Median Income Bracket | Number of ZIP Codes | Incidents | Probability Distance to Emergency Room ≤ 1 mi |
|---|---|---|---|
| $25,000 - $50,000 | 6 | 808 | 0.115 |
| $50,000 - $75,000 | 15 | 2120 | 0.759 |
| $75,000 - $100,000 | 7 | 915 | 0.905 |
| $100,000 - $200,000 | 9 | 889 | 0.761 |

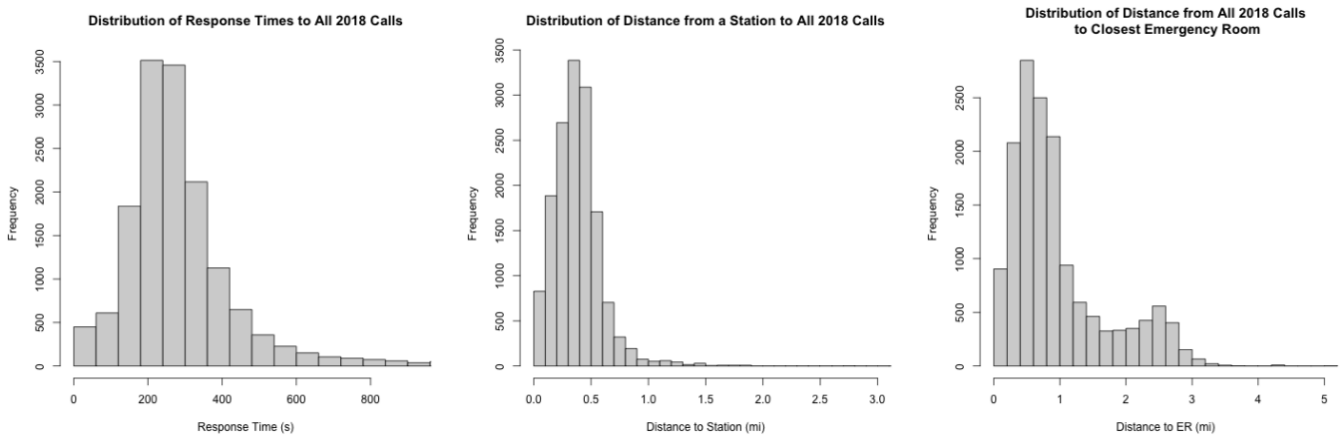

Figure 4 - (left to right) (a) Histogram of 2018 SFFD response times, (b) Histogram of 2018 distances from SFFD stations, (c) Histogram of distances to emergency rooms

## 4. Conclusions

Part of the responsibilities of the SFFD is to provide first responders across emergencies, not just fires, which account for only 12% of calls. On average, emergency calls were within 0.4 miles from a station, and firefighters were able to respond within 5 ½ minutes. However, we found ZIP codes with the lowest median incomes ($25,000 - $50,000) had the lowest probabilities to meet the average SFFD response threshold whether in time or distance—57.1% and 48.2% respectively, in 2018. The probability of medical emergencies in these low income ZIP codes being within the city's average distance to an emergency room was also significantly less than that of any higher income ZIP codes—at 11.5% in 2018.

There are several limitations to this study. We only examined the relationship of an area's median income bracket with the response times and distances to emergency services via simple logistic regression, and did not control for any potentially confounding factors. While the dataset spanned 2003 through 2021, analysis that accounts for changes in the distribution over time could provide value. Additionally, income data by ZIP code more recent than 2018 could change the estimate of equity if they reflect a change in income distribution.

As San Francisco considers equity as a part of California's "Blueprint for a Safer Economy," it should evaluate the distribution of access to emergency services. In particular, the potential to support citizens in lower income parts of the city with another emergency room provides opportunity.

## 5. Funding

This research did not receive any specific grant from funding agencies in the public, commercial, or not-for-profit sectors.

## 6. Declaration of Competing Interests

None declared.

**Appendix A - Average Responses from Each ZIP Code**

| ZIP Code | Average Time for Response | Average Distance from Station | Average Distance to Emergency Room | Median Income Bracket | Incidents | Population | Total Estimated Property Losses | Per Capita Property Losses |
|---|---|---|---|---|---|---|---|---|
| 94014 | 00:11:03 | 0.712 | 1.969 | $25-50k | 10 | 43810 | 0 | 0.000 |
| 94112 | 00:05:56 | 0.466 | 2.578 | $25-50k | 12523 | 77090 | 33639417 | 436.366 |
| 94124 | 00:05:55 | 0.391 | 1.622 | $25-50k | 15104 | 35550 | 51467239 | 1447.742 |
| 94130 | 00:07:08 | 0.502 | 2.243 | $25-50k | 1355 | 1690 | 2880469 | 1704.420 |
| 94134 | 00:06:18 | 0.485 | 2.319 | $25-50k | 5439 | 40590 | 12561765 | 309.479 |
| 94601 | 00:14:10 | 1.818 | 4.275 | $25-50k | 1 | 46310 | 0 | 0.000 |
| 94102 | 00:04:59 | 0.342 | 0.579 | $50-75k | 17100 | 23120 | 22268297 | 963.162 |
| 94103 | 00:05:30 | 0.359 | 0.934 | $50-75k | 14845 | 24670 | 14694293 | 595.634 |
| 94108 | 00:05:12 | 0.304 | 0.302 | $50-75k | 3572 | 10240 | 3295508 | 321.827 |
| 94110 | 00:05:13 | 0.386 | 0.925 | $50-75k | 16990 | 60330 | 55539874 | 920.601 |
| 94116 | 00:05:57 | 0.537 | 1.998 | $50-75k | 5389 | 40680 | 8919113 | 219.251 |
| 94121 | 00:05:47 | 0.393 | 1.947 | $50-75k | 8235 | 37230 | 17908850 | 481.033 |
| 94122 | 00:05:47 | 0.448 | 1.682 | $50-75k | 10425 | 50980 | 29671286 | 582.018 |
| 94132 | 00:06:13 | 0.631 | 2.649 | $50-75k | 6218 | 21800 | 4439350 | 203.640 |
| 94133 | 00:05:12 | 0.294 | 0.578 | $50-75k | 7211 | 21190 | 41790926 | 1972.200 |
| 94501 | 00:17:23 | 0.015 | 0.626 | $50-75k | 1 | 57830 | 0 | 0.000 |
| 94901 | 00:04:18 | 0.855 | 1.294 | $50-75k | 1 | 37280 | 0 | 0.000 |
| 94005 | 00:07:40 | 0.553 | 2.501 | $75-100k | 131 | 4640 | 123501 | 26.617 |
| 94109 | 00:04:55 | 0.293 | 0.350 | $75-100k | 19243 | 44290 | 37692952 | 851.049 |
| 94115 | 00:04:50 | 0.274 | 0.465 | $75-100k | 12267 | 28090 | 29702799 | 1057.415 |
| 94117 | 00:04:58 | 0.331 | 0.549 | $75-100k | 8019 | 31340 | 25773372 | 822.379 |
| 94118 | 00:05:16 | 0.433 | 0.815 | $75-100k | 5757 | 32970 | 33647409 | 1020.546 |
| 94129 | 00:06:52 | 0.714 | 2.089 | $75-100k | 1307 | 3080 | 1813645 | 588.846 |
| 94611 | 00:04:00 | 0.323 | 0.316 | $75-100k | 1 | 36730 | 0 | 0.000 |
| 94104 | 00:05:35 | 0.330 | 0.528 | $100-200k | 2795 | 1860 | 2451875 | 1318.212 |
| 94105 | 00:05:33 | 0.375 | 0.862 | $100-200k | 9735 | 10830 | 12177408 | 1124.414 |
| 94107 | 00:05:35 | 0.313 | 0.701 | $100-200k | 9900 | 26410 | 14917678 | 564.850 |
| 94111 | 00:05:06 | 0.265 | 0.588 | $100-200k | 3609 | 4210 | 11960601 | 2840.998 |
| 94114 | 00:05:08 | 0.327 | 1.168 | $100-200k | 8713 | 29310 | 30150794 | 1028.686 |

| 94123 | 00:05:36 | 0.394 | 1.154 | $100-200k | 6384 | 20650 | 9919396 | 480.358 |
| 94127 | 00:06:10 | 0.457 | 1.762 | $100-200k | 2853 | 19430 | 11123801 | 572.506 |
| 94131 | 00:06:05 | 0.360 | 1.472 | $100-200k | 3956 | 25720 | 10424499 | 405.307 |